\documentstyle[12pt]{article}
\begin{document}
\input epsf 
\newcommand{\Od}{{\cal O}}
\newcommand{\lsim}   {\mathrel{\mathop{\kern 0pt \rlap
  {\raise.2ex\hbox{$<$}}}
  \lower.9ex\hbox{\kern-.190em $\sim$}}}
\newcommand{\gsim}   {\mathrel{\mathop{\kern 0pt \rlap
  {\raise.2ex\hbox{$>$}}}
  \lower.9ex\hbox{\kern-.190em $\sim$}}}
%\vspace*{5mm}
\begin{center}
%\Huge{Essay on Gravitation}\\
%\LARGE{ESSAY ON GRAVITATION} \\
\vspace{2cm}
\Large{\bf Dark Geometry$^*$} \\
%\Large{\bf A Geometric Dark Matter Candidate} \\
\vspace*{1cm} 
\large{\bf J.A.R. Cembranos, A. Dobado and 
A.L. Maroto}  \\
\vspace{0.3cm}
\normalsize 
Departamento de F\'{\i}sica Te\'orica\\
Universidad Complutense de Madrid\\
28040 Madrid, Spain

\vspace*{1cm}  
{\bf ABSTRACT}\\  
\end{center}

Extra-dimensional theories contain additional degrees
of freedom related to the geometry of the extra space which
can be interpreted as new particles. Such theories allow to
reformulate most of the fundamental problems of physics 
from a completely different point of view.   
In this essay we concentrate on the brane fluctuations which are
present in brane-worlds, and how such oscillations of the own
space-time geometry along curved extra dimensions can help to
resolve the Universe missing mass problem.
The energy scales involved in these models are low compared
to the Planck scale, and this means that some 
of the brane fluctuations distinctive signals could be detected
in future colliders and in direct or indirect dark matter searches.

%\vspace*{2cm}
%\noindent
%$^1$ cembra@fis.ucm.es\\
%$^2$ dobado@fis.ucm.es\\
%$^3$ maroto@fis.ucm.es

\vspace*{0.5cm}
\noindent 
\rule[.1in]{8cm}{.002in}

\noindent 
$^*$ Essay selected for "Honorable Mention" in the 2004 Awards for 
Essays on Gravitation (Gravity Research Foundation). 
\newpage
\baselineskip 20pt
The elegant geometric description of the gravitational interaction 
which is the heart of General Relativity (GR) has  inspired 
several attempts
to find higher-dimensional generalizations  which
could unify all the fundamental interactions in a single 
picture. Based on  the early proposal by Kaluza and Klein 
(KK) \cite{KK}, the space-time was assumed
to have more than four dimensions, and the extra spatial dimensions
were compactified with a  tiny radius of the order of the 
Planck length. 
Thus, the momentum 
component along  the extra dimensions was seen from
the four-dimensional viewpoint as the mass of the KK
tower states associated to the ordinary particles. In this sense a 
new infinite set of 
particles arose whose mass spectrum was determined by the 
geometry of the extra space.
This would be the first example of a theory 
in which, not only the geometry of space-time is fixed by its matter
content as in GR, but also the properties of the 
own matter fields are determined by the
space-time geometry. This appealing possibility was also 
present somehow in the old string theories, whose consistency 
required
the introduction of six additional dimensions which typically
were also compactified at the Planck scale. Here again the 
topology and the geometry of the extra dimension space determined
the particle content. 

More recently, the construction of extra-dimensional models has been
revived within  the so called brane-world scenario \cite{ADD}.
These models are inspired in the conjectured M-theory which pretends to
be an unification of the old known consistent string theories 
through a web of different 
dualities.  M-theory includes also non-perturbative
effects leading to the emergency of a new extra dimension 
and a complete set of new higher-dimensional states generically 
called branes.

Unlike the old Kaluza-Klein theories, in the new brane-world
models the size
of the extra dimensions could be as large as a fraction
of a millimeter. The main assumption of this scenario is that 
by some (unknown) mechanism, matter fields are constrained to live
in a three-dimensional hypersurface (brane) embedded in the higher
dimensional (bulk) space. Only gravity is able to propagate in 
the bulk space, but the fundamental scale of gravity in $D$ 
dimensions
$M_D$ can be much lower than the Planck scale, the volume
of the extra dimensions being responsible for the actual value
of the Newton constant in four dimensions.

Since in these models gravity can live in the bulk, 
a tower of KK modes would be present in the theory, but
in addition,
new fields which are characteristic (distinctive) of these models 
and also have
a strong geometric origin (branons) appear in our
four dimensional space-time. In this essay we will
concentrate on some interesting properties of these fields.
Indeed, since  the energy scales involved in the brane-world
scenario can be much lower than the Planck scale, 
 the phenomenological consequences of branon physics
could be found in current collider experiments or in
astrophysical and cosmological observations.  

The fact that rigid objects are incompatible with GR
 implies
that the brane-world must be  dynamical and can move and
fluctuate along the extra dimensions. Branons are
precisely the fields parametrizing the
position of the brane in the extra coordinates.
Thus, in four dimensions 
branons could be
detected through their contribution to the 
induced space-time metric.
From a more fundamental point of view, branons can be understood as the
Goldstone bosons (GB) 
corresponding to the spontaneous breaking of the traslational
invariance along the extra dimensions which is produced by the 
presence of the brane \cite{Sundrum}. 

Let us consider  our four-dimensional space-time 
$M_4$ to be
embedded in a $D$-dimensional bulk space whose coordinates 
will be denoted by $(x^{\mu},y^m)$, where 
$x^\mu$, with $\mu=0,1,2,3$, correspond to the 
ordinary four dimensional space-time and $y^m$, with 
$m=4,5,\dots,D-1$, are coordinates of the compact extra
space.
For simplicity we will assume that the bulk 
metric tensor takes the following form:
\begin{eqnarray}
ds^2=\tilde g_{\mu\nu}(x)W(y)dx^\mu dx^\nu- g'_{mn}(y)dy^m dy^n
\label{metric}
\end{eqnarray}
where the warp factor is normalized as $W(0)=1$.

Working in the probe-brane approximation, our
3-brane universe is moving in the background metric given 
by (\ref{metric})
which is not perturbed by its presence. 
The position of the brane in the bulk can be parametrized as
$Y^M=(x^\mu, Y^m(x))$, and  we assume for simplicity that the ground
state of the brane corresponds to $Y^m(x)=0$. 

In the simplest case in which the metric is not
warped along the extra dimensions, 
i.e. $W(y)=1$,  
the transverse brane fluctuations are massless and
they can be parametrized by the GB fields 
$\pi^\alpha(x),\; \alpha=4,5, \dots D-1$.  
 In that case we can choose the $y$
coordinates   so that the branon fields are
proportional to the extra-space coordinates:
$\pi^\alpha(x)
=f^2\delta_m^\alpha Y^m(x)$ ,
where the proportionality constant is related to the brane 
tension $\tau=f^4$.

In the general case, the curvature generated by the
warp factor explicitly breaks the traslational 
invariance in the extra space. Therefore branons  acquire a mass 
matrix which is given precisely by the bulk Riemann tensor
evaluated at the brane position:
\begin{eqnarray}
M^2_{\alpha\beta}=\tilde g^{\mu\nu}R_{\mu\alpha\nu\beta}\vert_{y=0}
\end{eqnarray}

The fact that the brane can fluctuate implies that the actual
metric on the brane is no longer given by 
$\tilde g_{\mu\nu}$, but by the induced metric which includes
the effect of warping through the mass matrix:
\begin{eqnarray}
g_{\mu\nu}(x,\pi)
=
\tilde g_{\mu\nu}(x)\left(1+\frac{M^2_{\alpha\beta}\pi^\alpha \pi^\beta}{4f^4}\right)
%\eta_{\mu\nu}+\frac{1}{2}\partial_m\partial_n
%\tilde g_{\mu\nu}(x,y)\vert_{y=0}\;\delta^m_\alpha\delta^n_\beta
%\frac{\pi^\alpha\pi^\beta}{f^4}
-\frac{1}{f^4}\partial_{\mu}\pi^\alpha
\partial_{\nu}\pi^\alpha +\Od(\pi^4), 
\end{eqnarray}

The dynamics of branons can be obtained from the 
 Nambu-Goto action by introducing the above expansion.
In addition, it is also possible to get their couplings to
the ordinary particles just by replacing the space-time by
the induced metric in the Standard Model (SM) action.
Thus we get up to quadratic terms in the branon fields: 
\begin{eqnarray}
S_{Br} 
&=&\int_{M_4}d^4x\sqrt{\tilde g}\left[\frac{1}{2}
\left(\tilde g^{\mu\nu}\partial_{\mu}\pi^\alpha
\partial_{\nu}\pi^\alpha
-M^2_{\alpha\beta}\pi^\alpha \pi^\beta
\right)\right.\nonumber \\
&+&\left.\frac{1}{8f^4}\left(4\partial_{\mu}\pi^\alpha
\partial_{\nu}\pi^\alpha-M^2_{\alpha\beta}\pi^\alpha \pi^\beta 
\tilde g_{\mu\nu}\right)
T^{\mu\nu}_{SM}\right]
\nonumber \\
\label{Nambu}
\end{eqnarray}

We can see that branons  interact with the  SM
 particles through their energy-momentum tensor.
The couplings are controlled by the brane 
tension scale $f$ and they are  universal very much like
those of gravitons. For large $f$, branons are therefore 
weakly interacting particles.

The sign of the branon fields is determined by the orientation
of the brane submanifold in the bulk space. Under a parity 
transformation on the brane ($x^i\rightarrow -x^i$), the orientation
of the brane changes sign provided the ordinary space has an odd number
of dimensions, whereas it remains unchanged for even spatial dimensions.
In the case in which we are interested with three ordinary spatial
dimensions, branons are therefore pseudoscalar particles.
Parity on the brane then requires that branons always couple 
to SM particles by pairs, which ensures that they are 
stable particles. This fact can have important consequences
in cosmology as we show below.

Recent  observations \cite{WMAP} seem to confirm
that the universe contains an important fraction of non-luminous 
matter $\Omega_{DM}\sim 0.23$. Moreover, it is known that 
most of that matter 
cannot be made of any known
particle. Thus, the existence of dark matter candidates beyond
the SM or appropriate modifications of GR appear as 
inescapable consequences of the problem. 

In an expanding universe the number density of a massive, stable 
particle species
follows the thermal equilibrium abundance and
declines exponentially with the temperature, provided it is in 
equilibrium with radiation. However, if the species decouples, 
its abundance remains frozen with respect to the entropy density.
This means that if decoupling occurs early enough, i.e. the species is 
weakly interacting, its relic
number density could be cosmologically relevant and explain the
missing mass problem. Accordingly, we find that,
 in the brane-world scenario, the fluctuations of the 
own space-time geometry
along  curved extra dimensions are natural dark matter candidates, 
since as seen before, 
they satisfy all these requirements \cite{CDM}.

\begin{figure}[h]
%\epsfxsize=10cm   %width of figure - will enlarge/reduce the figures
%\epsfbox{fig3.eps}
%\figurebox{2cm}{3cm}{} %to have a box alone 
\centerline{\epsfxsize=12cm\epsfbox{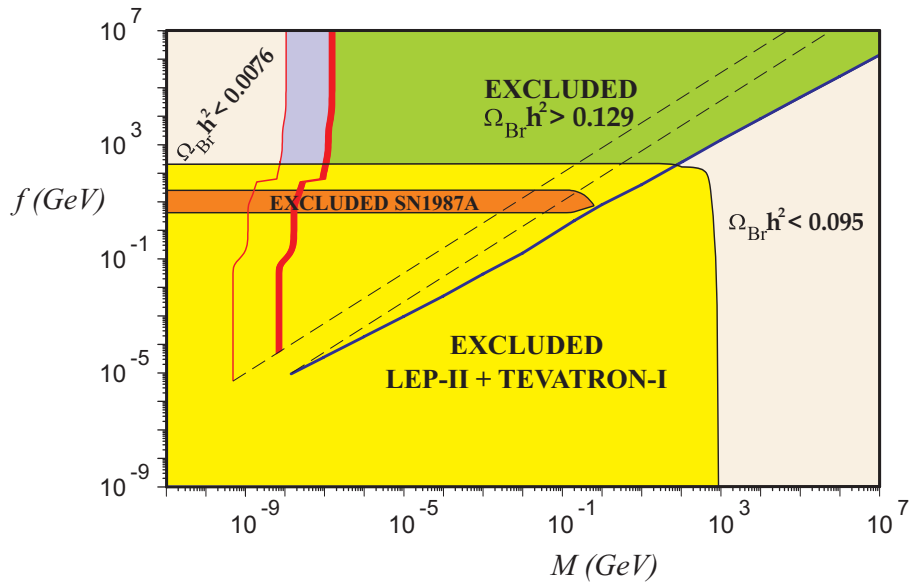}}
\caption{\footnotesize Relic abundance in the $f-M$ plane for a model with one branon of
mass: $M$. The two (red) lines on the left correspond to the
WMAP limits  \cite{WMAP} 
$\Omega_{Br}h^2=0.0076$ and 
$\Omega_{Br}h^2=0.129 - 0.095$  for hot-warm relics, whereas the
(blue) right line
corresponds to the latter limit for cold relics. 
The lower area is excluded by single-photon processes at LEP-II 
 together 
with monojet signal at Tevatron-I. The upper area is also excluded by 
cosmological branon overproduction. The astrophysical constraints are less
restrictive and they mainly come from supernova cooling by branon emission.}
\end{figure}
%\eject   

An explicit computation shows that
for sufficiently large $f$ and $M$, branons can 
constitute the total  dark matter of the universe (see Fig. 1). In
particular, they would contribute as cold dark matter 
in the  parameter range:   $M\sim f\gsim 200$ GeV.
The large value of $f$ would explain why they have not been detected
yet in collider experiments.
Another interesting possibility is the case with very light branons: 
$M\sim 100$ eV and $f\gsim 200$ GeV, where they would play
the role of  hot dark matter (Fig. 1).
 
Branons could be, not only natural cosmological dark matter 
candidates, 
but they could also make up the galactic halo and explain the 
local dynamics. In such a case, they could be detected in 
future direct
search experiments such as CDMS, CRESST II or GENIUS through their 
interactions with target nucleons. 
They could also been found in indirect
searches by the MAGIC or GLAST telescopes due to 
their annihilation into photons in the galactic halo; or 
by antimatter 
detectors such as AMS because of their
annihilation into charged particles. 
%Another possibility is the annihilation into neutrinos
%in the center of the Sun or the Earth, which could be observed by 
%high-energy neutrino telescopes such as AMANDA, IceCube or ANTARES.
These searches complement those in future high-energy colliders, 
such as
LHC or Linear Colliders which could find branon signals 
provided $f\lsim 1$ TeV
and $M\lsim 6$ TeV. If this possibility were realized in Nature,
it would provide a fascinating link between the geometry of our
space-time and an important fraction of the matter content of the
universe, providing a solution to the dark matter problem.

\newpage

 {\bf Acknowledgements:} 
 This work
 has been partially supported by the DGICYT (Spain) under the
 project numbers FPA 2000-0956 and BFM2002-01003.

\end{document}